\documentclass[10pt,conference]{IEEEtran}
\IEEEoverridecommandlockouts

\usepackage{graphicx}
\usepackage{amsmath,amssymb}
\usepackage{booktabs}
\usepackage{cite}
\usepackage{tabularx}
\usepackage{url}
\usepackage{booktabs,tabularx,ragged2e,threeparttable}
\usepackage{stfloats} 
\newcolumntype{Y}{>{\RaggedRight\arraybackslash}X}
\usepackage{array}
\usepackage{makecell}
\newcolumntype{L}[1]{>{\RaggedRight\arraybackslash}p{#1}}

\begin{document}

\title{AutoBinder Agent: An MCP-Based Agent for End-to-End Protein Binder Design}

\author{
\IEEEauthorblockN{
Fukang Ge\textsuperscript{1,†},
Jiarui Zhu\textsuperscript{1,†},
Linjie Zhang\textsuperscript{2},
Haowen Xiao\textsuperscript{1},
Xiangcheng Bao\textsuperscript{3},\\[-1pt]
Fangnan Xie\textsuperscript{1},
Danyang Chen\textsuperscript{1},
Yanrui Lu\textsuperscript{3},
Yuting Wang\textsuperscript{1},
Ziqian Guan\textsuperscript{1},\\[-1pt]
Lin Gu\textsuperscript{4},
Jinhao Bi\textsuperscript{5,*},
Yingying Zhu\textsuperscript{1,*}
}

\IEEEauthorblockA{
\textsuperscript{1}Guangzhou Institutes of Biomedicine and Health, Chinese Academy of Sciences\\[2pt]
\textsuperscript{2}Renmin University of China, 
\textsuperscript{3}South China University of Technology\\[2pt]
\textsuperscript{4}RIKEN, Japan, 
\textsuperscript{5}Westlake University
}

\IEEEauthorblockA{
\textsuperscript{†}\,These authors contributed equally to this work.\quad
\textsuperscript{*}\,Co-corresponding authors. \\
Contact: \texttt{zhu\_yingying@gibh.ac.cn}, \texttt{bijinhao@westlake.edu.cn}
}
}

\maketitle

\begin{abstract}
Modern AI technologies for drug discovery are distributed across heterogeneous platforms—including web applications, desktop environments, and code libraries—leading to fragmented workflows, inconsistent interfaces, and high integration overhead. We present an agentic end-to-end drug design framework that leverages a Large Language Model (LLM) in conjunction with the Model Context Protocol (MCP) to dynamically coordinate access to biochemical databases, modular toolchains, and task-specific AI models. The system integrates four state-of-the-art components: MaSIF (MaSIF-site and MaSIF-seed-search) for geometric deep learning-based identification of protein–protein interaction (PPI) sites, Rosetta for grafting protein fragments onto protein backbones to form mini proteins, ProteinMPNN for amino acid sequences redesign, and AlphaFold3 for near-experimental accuracy in complex structure prediction. Starting from a target structure, the framework supports de novo binder generation via surface analysis, scaffold grafting and pose construction, sequence optimization, and structure prediction. Additionally, by replacing rigid, script-based workflows with a protocol-driven, LLM-coordinated architecture, the framework improves reproducibility, reduces manual overhead, and ensures extensibility, portability, and auditability across the entire drug design process.
\end{abstract}


\section{Introduction}

Traditional drug design is a complex and time-consuming process, often requiring years of optimization from target identification to lead validation~\cite{kola2004_attrition}. 
Despite advances in Structure-Based Drug Design (SBDD) and High-Throughput Screening (HTS), these strategies still face limitations such as low throughput, restricted chemical space, and high attrition rates~\cite{macarron2011_hts}. 
Accelerating molecular discovery while maintaining biological rationality thus remains a key challenge in modern biomedicine~\cite{shih2018drugdiscoveryeffectiveness}.

Recent advances in AI agents offer a novel approach to optimizing traditional drug discovery workflows. By leveraging large language models (LLMs) as central reasoning engines, these agents can autonomously execute complex tasks in an end-to-end manner. Building on the success of multi-agent systems that demonstrated autonomous molecular design, this work focuses on developing a fully automated, end-to-end protein binder design agent tailored for AI-driven drug discovery (AIDD).

Recent molecular design approaches integrate MaSIF for geometric deep learning-based identification of protein–protein interaction sites (MaSIF-site, MaSIF-seed-search)~\cite{gainza2020masif,marchand2023masifseed};Rosetta for fragment grafting and mini-protein construction~\cite{leman2020rosetta}; ProteinMPNN for full-sequence redesign and amino acid optimization~\cite{dauparas2022proteinmpnn};and AlphaFold3 for near-experimental complex structure prediction~\cite{alphafold3_2024}. Together, these tools enable the \textit{de novo} design of protein binders directly from target structures, and are complemented by diffusion-based backbone generation such as RFdiffusion~\cite{watson2023rfdiffusion}.

While powerful, current protein design tools remain fragmented and difficult to integrate without expert knowledge. Prior pipelines, such as “Mapping targetable sites on the human surfaceome for the design of novel binders”~\cite{matthews2024_surfaceome}, are useful but still rely on manual coordination. Fully automated, end-to-end systems are rare. Meanwhile, AI-assisted frameworks such as the Virtual Lab of AI Agents and the Frogent end-to-end drug design agent have shown the feasibility of autonomous agentic workflows in drug discovery, yet have not been adapted for protein binder generation~\cite{swanson2025_virtual_lab,pan2025_frogent}.

\begin{figure}[!t]
  \centering
  \includegraphics[width=0.99\linewidth]{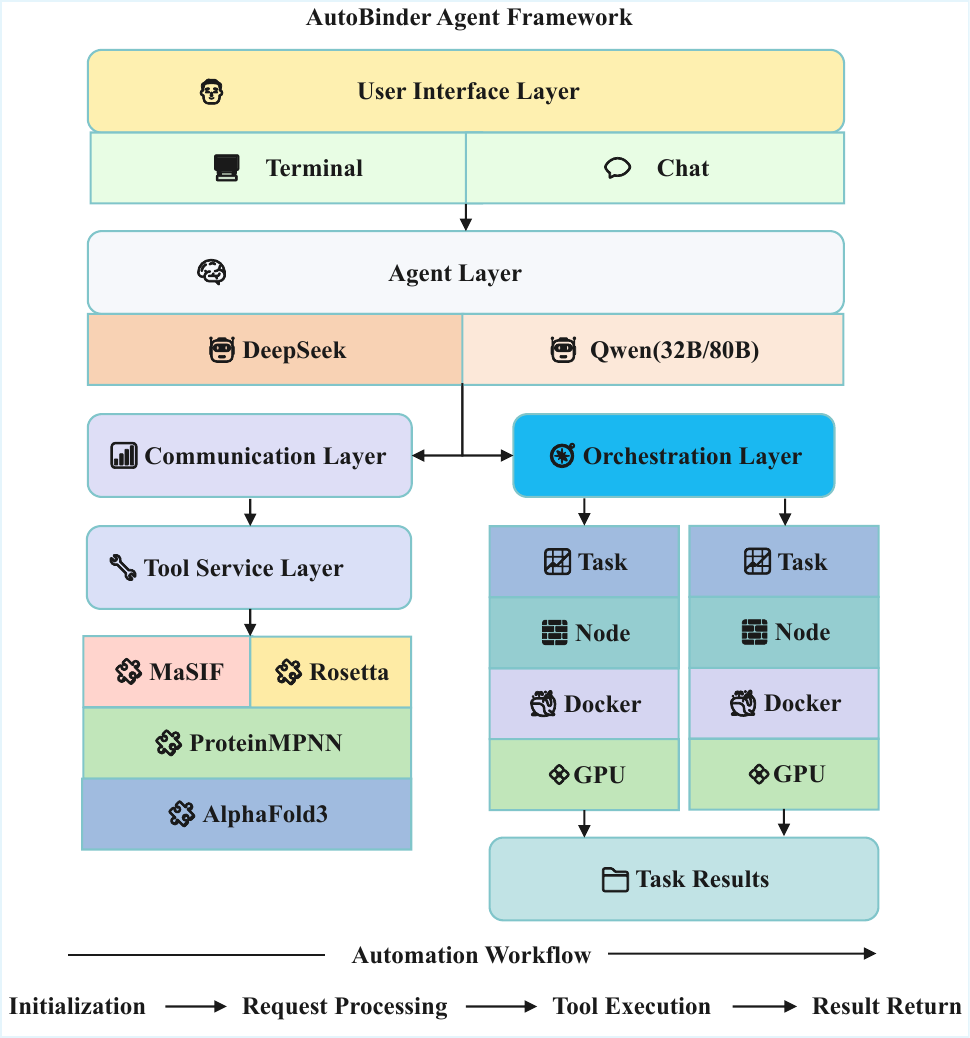}
  \caption{Framework of our proposed AutoBinder Agent}
  \label{fig:Frame}
\end{figure}

To bridge this gap, we present the AI-Driven Drug Design Agent (AutoBinder Agent), built on the Model Context Protocol (MCP). Unlike prior efforts that focus on model innovation, our work emphasizes system-level automation and standardization. AutoBinder Agent unifies MaSIF, Rosetta, ProteinMPNN, and AlphaFold3 into a reproducible, fully automated binder design workflow—from target input to candidate generation—while lowering the barrier for non-expert users.

Our key contributions include:

\begin{itemize}
\item \textbf{MCP-based integration:} AutoBinder Agent is the first to use MCP to standardize interaction among computational biology tools.
\item \textbf{End-to-end automation:} It transforms fragmented pipelines into a single, executable workflow.
\item \textbf{Practical validation:} We demonstrate its efficiency, scalability, and robustness in representative binder design tasks.
\end{itemize}

\section{Methods}

\subsection{System Architecture and Automation}

This study presents a multi-layer agentic system for protein design built upon the Model Context Protocol (MCP)~\cite{mcp_spec_2025}.
The architecture integrates large language model (LLM) reasoning, domain-specific computational tools, and distributed orchestration into an end-to-end automated workflow, bridging natural language interaction and scientific computation.

\vspace{3pt}
\noindent\textbf{System Architecture.}
As illustrated in Fig.~\ref{fig:Frame}, the framework comprises five hierarchical layers, each responsible for a distinct operational scope:

\begin{enumerate}
    \item User Interface Layer: Provides a command-line interface supporting human–AI collaborative interaction.
    \item Agent Layer: Implements heterogeneous LLM encapsulation via the MCP interface, supporting consistent function calling and contextual interaction.
    \item Communication Layer: Handles asynchronous message passing and context lifecycle management across processes.
    \item Orchestration Layer: Manages multi-node Docker container scheduling and GPU/NPU resource allocation for distributed computation.
    \item Tool Service Layer: Encapsulates protein design modules, including MaSIF, Rosetta, ProteinMPNN, and AlphaFold3.
\end{enumerate}

\vspace{3pt}
\noindent\textbf{Workflow.}
The automated protein design workflow comprises four sequential stages:
\begin{enumerate}
    \item Initialization: Load environment configurations, instantiate agents, and establish MCP communication sessions.
    \item Request Processing: The agent interprets user intent via LLM reasoning and generates structured tool invocation requests.
    \item Tool Execution: The tool service layer routes tasks to appropriate modules, performs computations, and serializes results.
    \item Resource Scheduling: The orchestrator dynamically allocates computational resources according to task priority and node status.
\end{enumerate}

\vspace{3pt}

\noindent\textbf{Model Integration and Automation.}
A unified adapter interface enables seamless integration of multiple LLMs. 
DeepSeek models are connected through OpenAI-compatible APIs, whereas the Qwen3 series are operated via local vLLM services. 
The scheduling mechanism employs priority queues and dynamic load balancing, 
and incorporates built-in exception capture, automatic retries, and resource recovery to ensure stable distributed execution.
\vspace{3pt}

\noindent\textbf{Tool Integration and Intelligent Scheduling.}
A key innovation of the system is its \textit{tool-integrated agentic workflow}.
The agent functions not only as a language model but also as an autonomous scheduler, 
capable of decomposing scientific objectives into sequential tool operations and orchestrating them automatically.
Through the human–computer interaction (HCI) interface, researchers can describe experimental goals in natural language, 
and the system autonomously generates reproducible computational workflows.

\subsection{Binder Design Workflow}
Given a user-specified target protein, the binder-design agent executes the following end-to-end workflow:

\begin{enumerate}
  \item Interaction-site prediction (maSIF-site): This stage should provide a detailed explanation of how maSIF-site performs interaction-site prediction.

  \item Seed retrieval and matching (maSIF-seed-search): For each interaction site predicted in Step~1, iterate over the seed-candidate database so that each seed is matched/docked to the site, and retain the best-scoring seed (protein fragment).

  \item Fragment grafting (Rosetta): Graft the selected protein fragments onto the protein backbone to obtain mini-protein candidates.

  \item Full-sequence redesign of the mini-protein (ProteinMPNN): Redesign all amino acids of the mini-protein; select the top 5 sequences ranked from low to high by global\_score.

  \item Structure prediction (AlphaFold3): Predict structures for the improved sequences and keep the top-ranked structure as the final binder (improved binder).
\end{enumerate}

\begin{figure}[!t]
  \centering
  \includegraphics[width=0.99\linewidth]{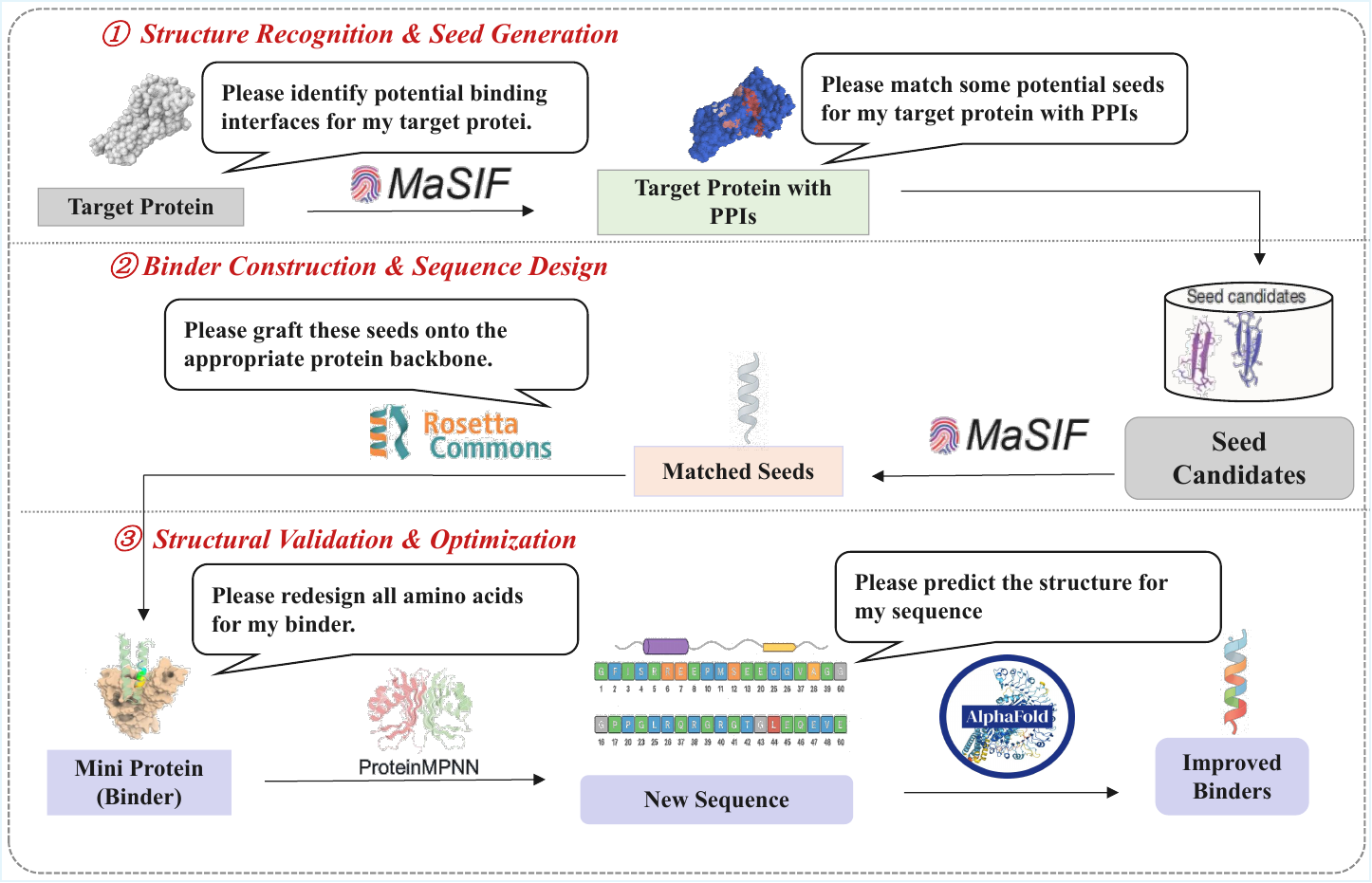}
  \caption{Workflow of Binder Design by Autobinder Agent}
  \label{fig:binder-workflow}
\end{figure}

\section{Experiments}

\subsection{Objective}

This experiment aims to validate whether the AutoBinder Agent, 
built on the Model Context Protocol (MCP) framework, 
can achieve fully automated protein design workflows from task parsing to structure generation under natural language instructions.  
The main focus is to assess the agent’s intelligent planning and execution performance 
in multi-tool cooperative scenarios.

The experiment focuses on the following five capability dimensions:

\begin{enumerate}
    \item \textbf{Task Understanding and Tool Planning:} Evaluate whether the agent can accurately identify task semantics and construct a reasonable tool invocation chain.
    \item \textbf{Workflow Consistency and Logical Order:} Verify whether the agent maintains correct dependency relationships and stable task sequencing.
    \item \textbf{Schema Parsing and Parameter Generation:} Examine whether the agent can transform natural language input into structured parameters compliant with MCP standards.
    \item \textbf{Error Recovery and Strategy Generation:} Assess whether the agent can generate fallback plans or retry mechanisms when encountering module exceptions.
    \item \textbf{End-to-End Design Verification:} Confirm whether the agent can autonomously complete the entire workflow from interaction-site prediction to binder generation.
\end{enumerate}

\subsection{Experimental Design}

To systematically evaluate the agent’s intelligent behavior in complex tasks, 
this experiment adopts a standardized prompt-driven testing protocol.  
Researchers issue natural language prompts to the system, 
and the agent, following the MCP framework, performs task parsing, tool orchestration, and result generation, 
forming a closed-loop computational workflow.

Each prompt corresponds to a specific core capability test, 
covering tool identification, workflow planning, parameter parsing, fault tolerance, and end-to-end execution.  
All tasks are executed in isolated contexts to avoid state interference.  
Table~\ref{tab:prompt_tasks} summarizes the five standardized prompt tasks and their verification objectives.

\begin{table}[t]
\centering
\caption{Standardized prompt tasks of the AutoBinder Agent}
\label{tab:prompt_tasks}
\footnotesize
\setlength{\tabcolsep}{4pt}
\renewcommand{\arraystretch}{1.12}
\begin{tabularx}{\columnwidth}{p{0.8cm} p{2.8cm} X}
\toprule
\textbf{ID} & \textbf{Capability Type} & \textbf{Prompt Example} \\
\midrule
\textbf{P1} & Tool Identification & ``List all available protein design tools under the MCP framework and their functions.'' \\
\textbf{P2} & Workflow Planning & ``Describe the complete automated process from an input PDB file to generating an improved binder.'' \\
\textbf{P3} & Parameter Parsing & ``Run maSIF-seed-search on PDB 4QVF\_A and set the number of matches to 20.'' \\
\textbf{P4} & Error Recovery & ``If AlphaFold3 is temporarily unavailable, how would you continue the task?'' \\
\textbf{P5} & End-to-End Design Verification & ``Based on the input target protein, complete the full design workflow from interaction-site prediction to binder generation.'' \\
\bottomrule
\end{tabularx}
\end{table}

\subsection{Evaluation Metrics}

To quantitatively evaluate the system behavior of the AutoBinder Agent, five behavioral metrics were defined.
These metrics are scored based on the agent’s responses to the standardized prompts, with each metric ranging from 0 to 100, where higher values indicate better performance.

\begin{itemize}
    \item \textbf{Task Understanding Accuracy} ($S_{\text{task}}$): Evaluates whether the agent correctly interprets the task semantics and maps them to appropriate workflow steps.
    \item \textbf{Tool Selection Accuracy} ($S_{\text{tool}}$): Measures whether the system selects and invokes the correct computational tools with logical dependency consistency.
    \item \textbf{Execution Order Consistency} ($S_{\text{order}}$): Assesses whether the sequence of tool invocations aligns with the expected protein design pipeline.
    \item \textbf{Reasoning and Interpretability} ($S_{\text{reason}}$): Determines whether the agent provides clear and logically consistent explanations for computational steps and MCP communication.
    \item \textbf{Output Consistency} ($S_{\text{consistency}}$): Tests the reproducibility and stability of outputs across repeated executions.
\end{itemize}

The overall composite score is computed as:

\begin{equation}
\begin{split}
S_{\text{prompt}} = {} & 0.30 S_{\text{task}} + 0.25 S_{\text{tool}} \\
                       & + 0.20 S_{\text{order}} + 0.15 S_{\text{reason}} 
                       + 0.10 S_{\text{consistency}}
\end{split}
\end{equation}

A task is considered validated when $S_{\text{prompt}} \ge 80$. 
Scores above 90 across multiple tasks indicate strong robustness and generalization in system-level behavior.

\section{Results and Analysis}

\subsection{Experimental Results}

\begin{figure}[!t]
  \centering
  \includegraphics[width=0.99\linewidth]{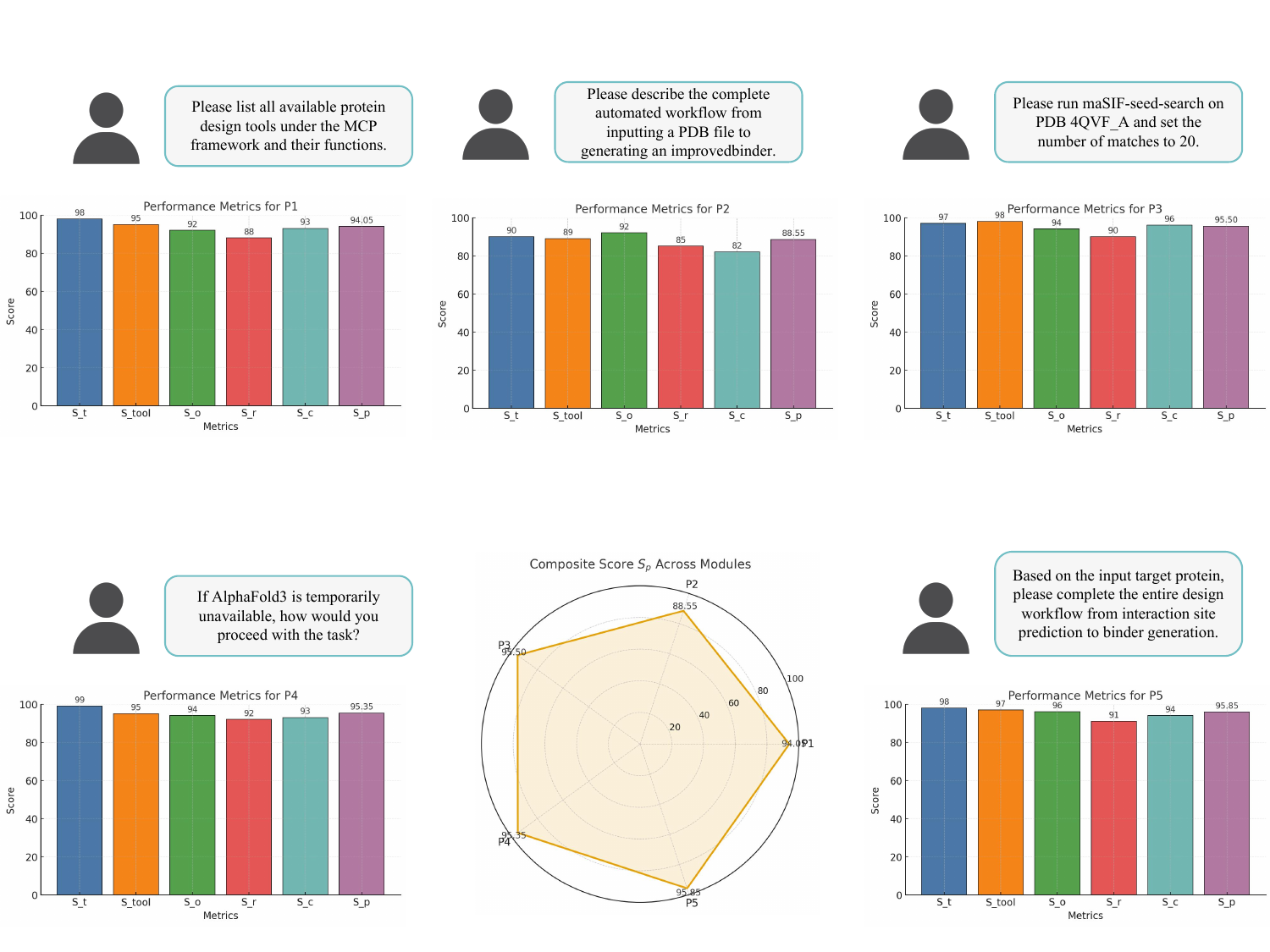}
  \caption{Results of All Experiments}
  \label{fig:Results}
\end{figure}

The AutoBinder Agent was evaluated on five standardized prompt tasks to assess its capabilities in 
task understanding, workflow planning, parameter parsing, error recovery, and end-to-end design execution. The results can be seen in Table~\ref{tab:results} and Fig.~\ref{fig:Results}.

Overall, the agent demonstrated high comprehension accuracy and stable automation behavior under the 
Model Context Protocol (MCP) framework.  
It correctly identified registered tools, constructed coherent task pipelines, 
and executed protein design workflows with minimal manual supervision.

The system achieved strong results in \textit{task understanding} and \textit{tool invocation}, 
demonstrating accurate interpretation of prompts and consistent mapping to appropriate MCP modules.  
Execution order remained logically coherent, and error recovery mechanisms successfully maintained workflow stability.  
However, schema validation details and output traceability were limited.

\begin{table}[!t]
\centering
\caption{Performance of AutoBinder Agent across standardized prompt tasks}
\label{tab:results}
\footnotesize
\setlength{\tabcolsep}{3.5pt}
\renewcommand{\arraystretch}{1.1}
\begin{tabular}{lcccccc}
\toprule
\textbf{Task} & $S_t$ & $S_{tool}$ & $S_o$ & $S_r$ & $S_c$ & $S_p$ \\
\midrule
P1 & 98 & 95 & 92 & 88 & 93 & 94.05 \\
P2 & 90 & 89 & 92 & 85 & 82 & 88.55 \\
P3 & 97 & 98 & 94 & 90 & 96 & 95.50 \\
P4 & 99 & 95 & 94 & 92 & 93 & 95.35 \\
P5 & 98 & 97 & 96 & 91 & 94 & 95.85 \\
\textbf{Avg.} & \textbf{96.4} & \textbf{94.8} & \textbf{93.6} & \textbf{89.2} & \textbf{91.6} & \textbf{93.86} \\
\bottomrule
\end{tabular}
\end{table}

\subsection{Analysis}

As shown in Table~\ref{tab:results}, the AutoBinder Agent achieves consistently high performance across all prompt tasks, with an overall score of $S_p = 93.86$. 
All metrics exceed the validation threshold, indicating strong robustness and stability. 
Notably, the high task understanding ($S_t = 96.4$) and tool selection ($S_{tool} = 94.8$) scores demonstrate a well-defined and effectively executed workflow, while execution order and consistency confirm reliable coordination.

These results primarily stem from the \textbf{deterministic workflow design}, 
where each tool in the MCP framework is bound to a fixed computational step. 
This clear structure minimizes ambiguity and ensures consistent, high-performing automation across tasks.

Overall, AutoBinder exhibits a stable and interpretable performance pattern, validating its capability for reliable automated protein design.

\section{Discussion}

AutoBinder effectively unifies multiple computational biology tools within an MCP-based framework, achieving automation and interpretability in protein binder design. 
The high overall performance confirms that the agent maintains a stable reasoning–execution loop across diverse tasks.

The results highlight the advantage of deterministic workflows—clear execution chains reduce ambiguity and enhance reproducibility. 
However, human-evaluated metrics such as reasoning clarity may introduce subjective bias. 
Future work will integrate more quantitative indicators to strengthen objectivity.

In summary, AutoBinder demonstrates the feasibility of an LLM-coordinated, protocol-driven framework for scientific computation and its promise for broader AI-driven molecular design applications.

\bibliographystyle{IEEEtran}
\bibliography{references}

\end{document}